\documentclass[prl,twocolumn,showpacs,preprintnumbers,amsmath,amssymb]{revtex4-1}
\usepackage{graphicx}
\usepackage{dcolumn}
\usepackage{bm}
\usepackage{color}
\usepackage[colorlinks,linkcolor=blue,urlcolor=blue,citecolor=blue]{hyperref}
\usepackage{CJK,CJKnumb,CJKulem}
\begin{document}
\begin{CJK*}{GBK}{com}

\title{Coupled Leidenfrost States as a Monodisperse Granular Clock}

\author{Rui Liu}
\email{lr@iphy.ac.cn}
\author{Mingcheng Yang}
\author{Ke Chen}
\email{kechen@iphy.ac.cn}
\author{Meiying Hou}
\email{mayhou@iphy.ac.cn}
\affiliation{Beijing National Laboratory for Condensed Matter Physics and CAS Key Laboratory of Soft Matter Physics, Institute of Physics, Chinese Academy of Sciences, Beijing 100190}
\author{Kiwing To}
\email{ericto@gate.sinica.edu.tw}
\affiliation{Institute of Physics, Academia Sinica, Taipei 115}

\date{\today}

\begin{abstract}
Using an event-driven molecular dynamics simulation, we show that simple monodisperse granular beads confined in coupled columns may oscillate as a new type of granular clock. To trigger this oscillation, the system needs to be driven against gravity into a density-inverted state, with a high-density clustering phase supported from below by a gas-like low-density phase (Leidenfrost effect) in each column. Our analysis reveals that the density-inverted structure and the relaxation dynamics between the phases can amplify any small asymmetry between the columns, and lead to a giant oscillation. The oscillation occurs only for an intermediate range of the coupling strength, and the corresponding phase diagram can be universally described with a characteristic height of the density-inverted structure. A minimal two-phase model is proposed and linear stability analysis shows that the triggering mechanism of the oscillation can be explained as a switchable two-parameter Hopf bifurcation. Numerical solutions of the model also reproduce similar oscillatory dynamics to the simulation results.
\end{abstract}
\pacs{45.70.Qj}

\maketitle
\end{CJK*}
Nonequilibrium systems may exhibit self-sustained oscillations, which play important roles in the generation of periodic rhythms in nature, especially in biological systems \cite{Pikovsky2001}. A purely physical and macroscopic illustration of such nonequilibrium oscillations is the recently discovered granular clock \cite{Lambiotte2005,Costantinia2005,Miao2005,Viridi2006,Hou2008,Liu2009,Li2012, Hussain2012}, which shows that bidiperse beads can oscillate horizontally between connected compartments under vertical vibrations. The oscillation is facilitated by the vertically heterogeneous species distribution of beads in each compartment and the coupling that naturally allows beads of different species to alternately move between the compartments. Unlike some previously-reported coupling induced oscillations \cite{In2003,In2006,Bulsara2004,Hernandez2008}, this oscillation does not rely on purposely-designed frustrations or unidirectional coupling mechanisms. However, the complicate conditions for the oscillation in this bidisperse system are too specific to provide any general ideas on how to make oscillations through simple couplings, which is a core problem in understanding many biological dynamics. Thus it would be of great interest to clearly show a coupling-induced spontaneous oscillation in an even simpler granular system.

In this Letter, we numerically show that density heterogeneity, instead of bidispersity, may drive a simple monodisperse granular system to oscillate in coupled columns, as a new type of granular clock. The coupling strength of the system can be monotonically tuned, and we find that the oscillation is triggered only for an intermediate range of the coupling strength. An oscillation phase diagram is mapped out for different total number of beads and vibrational strength.  A scaling relation for the phase diagram is found, which indicates that the occurrence of the oscillation depends critically on some structural matching relations of the system. Based on these results, a minimal two-phase model is proposed. Linear stability analysis and numerical solutions of the model qualitatively explain the triggering mechanism and the global dynamics of the oscillation.

\begin{figure}
\includegraphics[width=7.5 cm]{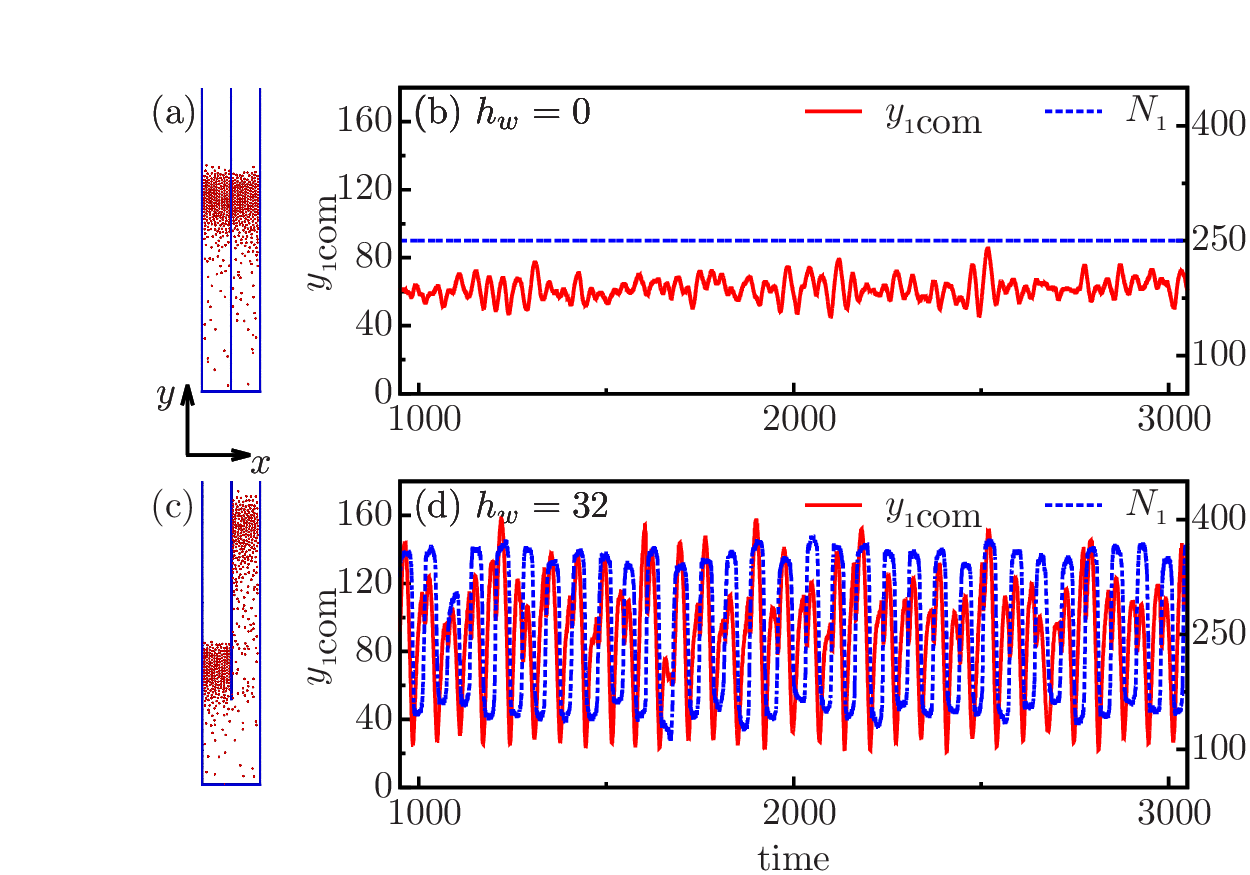}
\caption{\label{fig1} Simulation results for $N=500$ and $v_b=24$: (a) uncoupled granular columns with $h_w=0$; (b) the height of the center of mass ($y_{1\textrm{com}}$) and the number of beads ($N_1$) for the left column in (a); (c) two coupled granular columns with $h_w=32$; (d) $y_{1\textrm{com}}$ and $N_1$ for the left column in (c).}
\end{figure}

Using an event-driven molecular dynamics method \cite{Rapaport1997,Poeschel2005}, we simulate vertically vibrated two-dimensional columns of monodisperse spherical beads under gravity (in $y$-direction), as shown in Fig. \ref{fig1}. In our simulation, the mass $m$ and the diameter $d$ of the beads are both set to 1, and the time unit is $\sqrt{d/g}$ with $g$ being the gravitational acceleration. The width of each column is $W = 10$ \cite{arg-w}, and the height of the side-walls is considered to be infinite. By assuming a saw-tooth vibration with infinitely small amplitude $A$ and infinitely large frequency $f$, the bottom plate is assigned an constant upward velocity $v_b=Af$ but effectively kept stationary \cite{Eggers1999}. The coefficient of restitution between the beads is $e=0.9$, and the dissipation in a particle-wall collision is neglected. Typically a two-column system is simulated, and the columns are coupled through a bottom window with tunable height $h_w$ in the separation wall [Fig. \ref{fig1}(c)]. The particles are permitted to pass through the window if $h_w$ is large enough.

Initially, each column is filled with equal number of beads, namely $N_1(0)=N_2(0)=N/2$ (subscript from $1$ to $2$ denotes the column index from left to right) with $N=N_1+N_2$ being the total number. Each simulation runs with the window closed ($h_w=0$) at the beginning. For proper $N$ and $v_b$, each column may exhibit a Leidenfrost phenomenon \cite{Meerson2003,Eshuis2005}, i.e. the formation of a density-inverted structure with a high-density clustering phase (CP) supported from below by a gas-like low-density phase (GP) [Fig. \ref{fig1}(a)]. The floating cluster in an uncoupled column may fluctuate as a piston \cite{Brey2010} or show noisy resonances \cite{Rivas2013}. A similar irregular motion has also been observed in our simulation, as shown in Fig. 1(b). However, the irregular motion is relatively very small, and the system can still be thought to stay in a steady state. In this equally-partitioned steady state (EPSS), the two columns are statistically identical, and share the same vertical number-density profile $n^{0}(y)$ with the same characteristic height $h_{inv}$ of the floating cluster. Here, $n^{0}(y)$ is measured by counting the number of beads (averaged over time) per unit length in $y$-direction, and $h_{inv}$ is evaluated as the height corresponding to the maximum vertical gradient of $n^0(y)$ \cite{Eshuis2005}. After the EPSS is reached, the window is opened to the preset height of $h_w$. For two coupled columns with a proper $h_w$ [Fig. \ref{fig1}(c)-\ref{fig1}(d)], we find that the system oscillates fiercely with an amplitude several times larger than that of the irregular motion in the uncoupled case.

\begin{figure}
\includegraphics[width=7.5 cm]{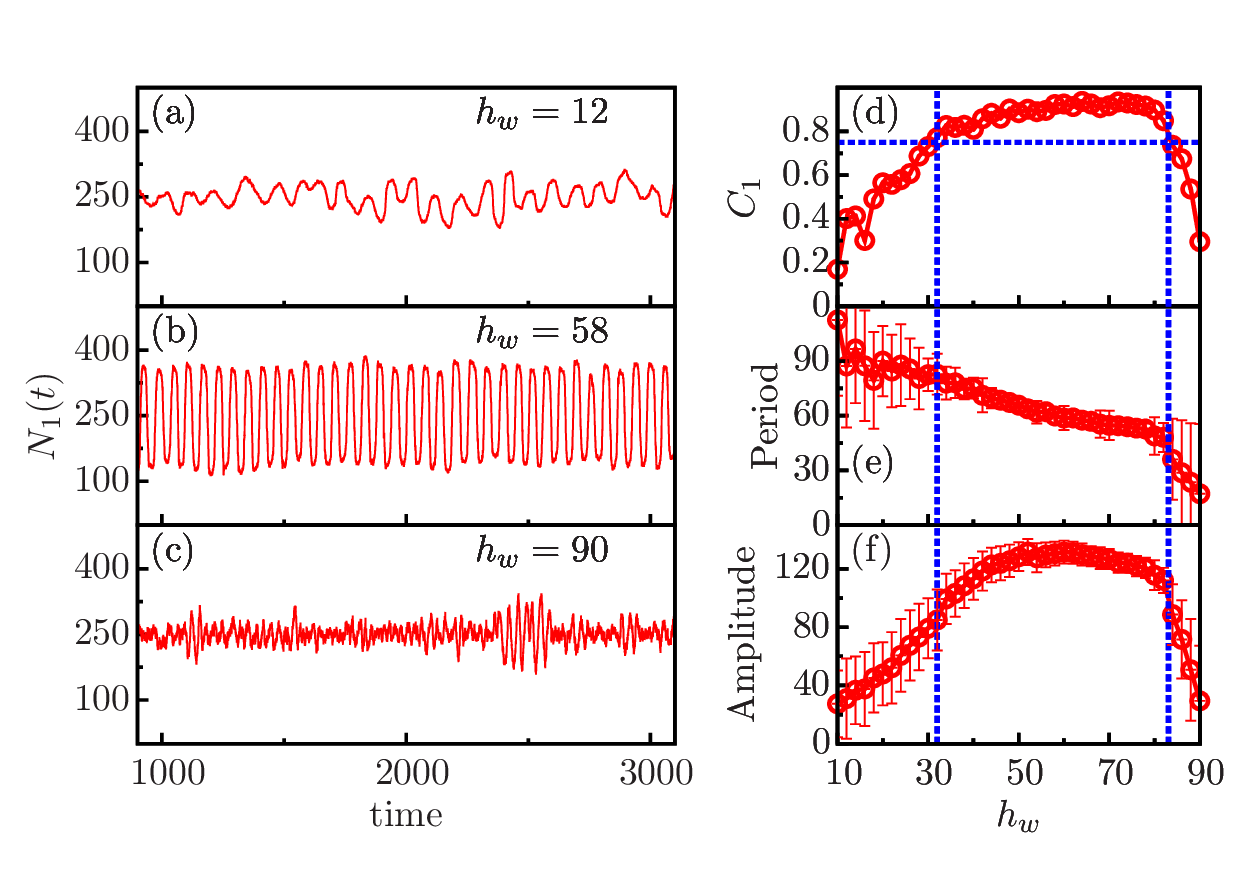}
\caption{\label{fig2} Coupled dynamics for $N=500$, $v_b=30$: $N_1(t)$ for weak coupling at $h_w=12$ (a), moderate coupling at $h_w=58$ (b), and strong coupling at $h_w=90$ (c); (d) the first positive peak value $C_1$ of the auto-correlation function of $s_1(t)$; the averaged period (e) and amplitude (f) of $s_1(t)$ for different $h$, respectively with relative error bars in arbitrary units.}
\end{figure}

For given $N$ and $v_b$, regular oscillations are observed only for an intermediate range of $h_w$. As illustrated in Fig. \ref{fig2}(a)-\ref{fig2}(c), a good oscillation is observed at $h_w=58$ for $N=500$ and $v_b=30$, but only a fluctuation-like behavior is observed at smaller $h_w=12$ or larger $h_w=90$. Generally, a larger $h_w$ indicates a stronger coupling between the columns, and the oscillation occurs only in the intermediate coupling regime. To investigate the coupling effect in triggering the oscillation, we have performed simulation runs with different $h_w$ for different $N$ and $v_b$. Each run results in a curve like that shown in either of Fig. \ref{fig2}(a)-\ref{fig2}(c). We distinguish the quality of the oscillation by calculating the auto-correlation function of $s_{i}(t) \equiv N_{i}(t)-N/2$ for each curve:
\begin{equation}
C(\tau) = \langle s_{i}(t)s_{i}(t+\tau)\rangle/\sigma^2,
\end{equation}
where $\sigma$ is the standard deviation of $s_i(t)$, $\tau$ is the time lag, and $\langle\cdot\rangle$ denotes a time average. $C(\tau)$ oscillates for perfect oscillations but quickly vanishes for pure fluctuations. We use the first positive peak value $C_1$ of $C(\tau > 0)$ to define the quality of an oscillation. For $N=500$ and $v_b=30$, $C_1$ as a function of $h_w$ is shown in Fig. \ref{fig2}(d). $C_1$ first increase and then decrease, with increasing $h_w$. At about $h_w=60$, $C_1$ reaches its maximum of almost 1 for a nearly perfect oscillation. Empirically, we define an acceptable oscillation with $C_1 \ge 0.75$ (above the dotted horizontal line). This gives a lower boundary $h_{LO}=32$ and a upper boundary $h_{HI}=82$ (dotted vertical lines) of the range of $h_w$ for acceptable oscillations.

Meanwhile, we perform Hilbert spectral analysis \cite{Pikovsky2001} on each oscillatory curve, which unambiguously gives the instantaneous phase $\phi(t) \in [0,2\pi)$ and amplitude $A(t)$ of the curve regardless of the oscillation quality:
\begin{equation}
s_{i}(t) + \mathbf{i}\mathcal{H}[s_{i}(t)] \equiv A_{i}(t)\exp[\mathbf{i}\phi_{i}(t)],
\end{equation}
where $\mathcal{H}[\cdot]$ denotes a Hilbert transform and $\mathbf{i}$ is the imaginary unit. Then, the apparent oscillation period can be measured from $\phi_i(t)$. For $N=500$ and $v_b=30$, the averaged period and amplitude are respectively shown in Fig. \ref{fig2}(e) and \ref{fig2}(f). In the regime $C_1 < 0.75$ ($h_w<h_{LO}$ or $h_w>h_{HI}$), which is assumed to be non-oscillatory, the relative errors for both the period and the amplitude are quite large. Actually, neither the period nor the amplitude is well-defined in this regime. In the oscillatory regime $C_1 \ge 0.75$ ($h_{LO} \le h_w \le h_{HI}$), the period decreases with increasing $h_w$, while the amplitude behaves similarly to $C_1$ and reaches a high plateau, which indicates giant oscillations.
\begin{figure}
\includegraphics[width=8 cm]{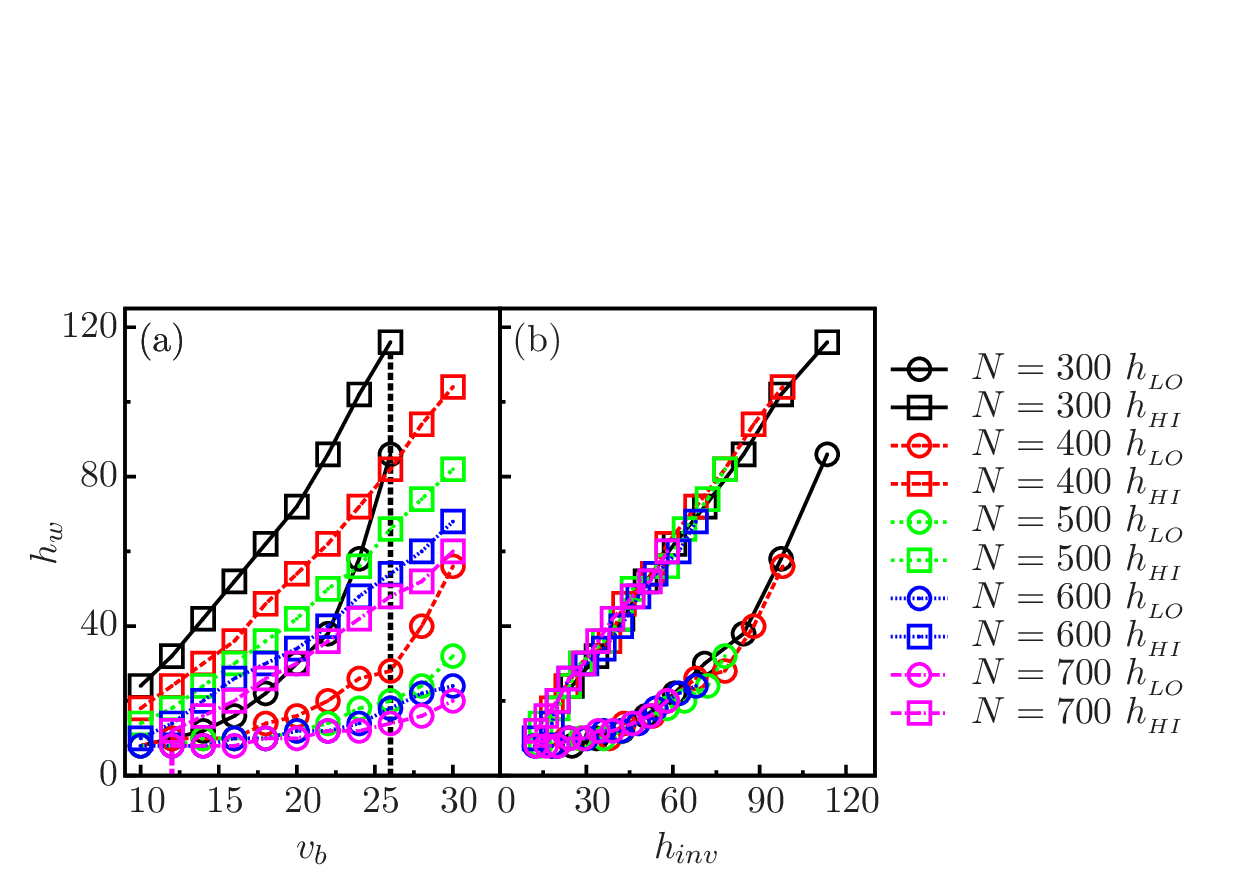}
\caption{
\label{fig3} Oscillation phase diagram: (a) oscillatory regimes with lower boundaries shown in empty circles and upper boundaries in empty squares, different $N$ is distinguished by colors, and no oscillation is observed for $N=700,\ v_b<12$ and $N=300,\ v_b>26$ (marked by vertical dotted lines); (b) Collapsed phase diagram in the $h_w$-$h_{inv}$ plane.
}
\end{figure}

\begin{figure}
\includegraphics[width=8cm]{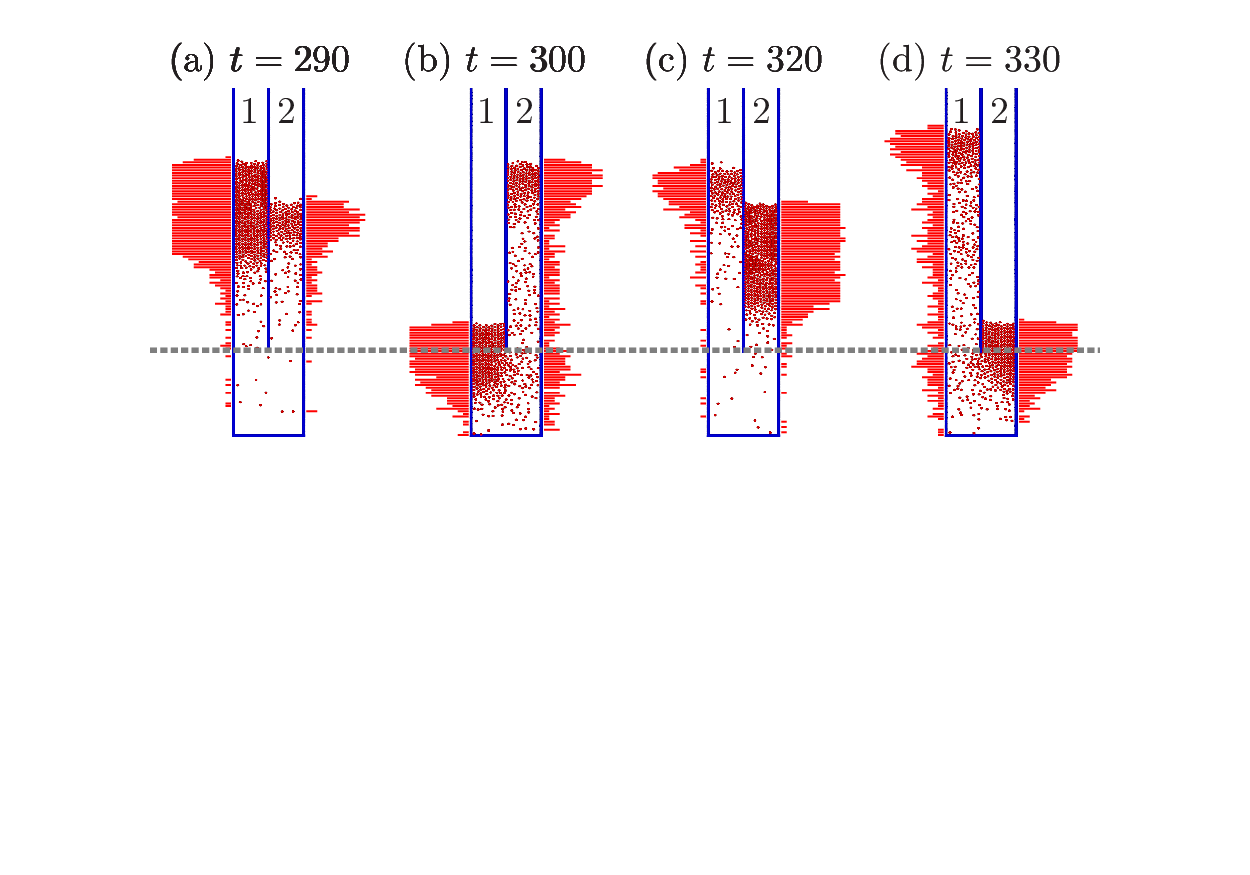}
\caption{
\label{fig4} Time-sequential snapshots with density analysis for one oscillation under $N=500$, $v_b=24$, and $h_w=32$: (a) $t=290$; (b) $t=300$; (c) $t=320$; (d) $t=330$, instantaneous density profiles are shown in horizontal bars on the left (column 1) or right (column 2) side of each granular column, the grey dashed line indicates the height of the window.
}
\end{figure}

With the empirical criterion $C_1 \ge 0.75$, an oscillation phase diagram can be obtained for different $N$ and $v_b$, as shown in Fig. \ref{fig3}(a). Obviously, both $h_{LO}$ and $h_{HI}$ increase with increasing $v_b$ but decrease with increasing $N$, just as $h_{inv}$ of the EPSS behaves. Then a simple idea is to replot the phase diagram for different $N$ and $v_b$ with a single parameter, $h_{inv}(N,v_b)$. Indeed, we obtain a collapsed phase diagram with $h_{inv}$ for different $N$ and $v_b$, as shown in Fig. \ref{fig3}(b). Thus, systems of different $N$ and $v_b$ but with the same $h_{inv}$ would share the same oscillatory regime $h_{LO}<h_w<h_{HI}$. The fact that $h_{LO}$ and $h_{HI}$ are only functions of $h_{inv}$, suggests that a structural matching between the window ($h_w$) and the initial density-inverted structure ($h_{inv}$) may play an important role in the oscillatory phenomenon. Two deductions on this structural matching concept can be made. First, when $h_w \ge h_{inv}$, the two columns share the same gaseous part, and actually merge into a wider non-oscillatory single-column. This explains the collapsed linear relation that $h_{HI} \approx h_{inv}$ in Fig. \ref{fig3}(b). Second, only if $h_w$ corresponds to the height of a large enough vertical gradient of $n^0(y)$, the triggering of the oscillation becomes possible (explained in Fig. \ref{fig4} and its context). As $n^0(y)$ can be roughly determined by $h_{inv}$ \cite{arg-hlo}, the collapse of $h_{LO}$ in the $h_w$-$h_{inv}$ plane can be understood. Further side support for the above concept is that no oscillation has been observed for $N=700,\ v_b<12$ or $N=300,\ v_b>26$. For large $N=700$ and small $v_b<12$, $n^0(y)$ is highly compressed due to the strong dissipation. Thus the range from $h_{LO}$ to $h_{HI}$ would be too small to be observed in our simulation. For small $N=300$ and large $v_b>26$, the whole system including the floating clusters tends to be gasified. The vertical gradient of $n^0(y)$ would be rather small, even if a density inversion is still present. As we have mentioned above, a small density gradient will not help in triggering an oscillation.

To verify the triggering mechanism of the oscillation, we inspect in detail the oscillatory process. Four time-sequential snapshots of one complete oscillation under $N=500,\;v_b=24$ and $h_w=32$ are shown in Fig. \ref{fig4}. At time $t=290$, a larger floating cluster is formed in column 1 than that in column 2 because of the excessive population ($N_1 > N_2$). The cluster in column 1 sinks into the window region due to the lack of enough support at $t=300$, and there the distinct density difference between the two columns drives a massive flow of beads from column 1 to 2. The beads that have entered column 2 are heated up by the bottom plate, and push the smaller cluster on top to a higher place. As dissipation increase dramatically with the increased population, the beads in column 2 start to condensate and form an even larger cluster at $t=320$. Meanwhile, a much smaller but higher cluster is formed in column 1 through an evaporation process due to the decreased population. The relaxation time needed by both the evaporation and condensation processes, as well as the large horizontal density difference, allows enough beads to transfer, which maintains a non-damping oscillation. After the evaporation and condensation, the situation becomes similar to that of Fig. \ref{fig4}(a), except that the two columns have been playing reversed roles. Then following the same process, column 2 drives beads back into 1, as shown in Fig. \ref{fig4}(d). The above oscillation picture is valid even in the triggering moment, when the two columns are almost identical. Large vertical gradient in $n^0(y)$ around $h_w$ may cause large horizontal density difference in the window region under perturbations. Thus any small population difference between the columns may be amplified through the above process and the oscillation can be triggered.
\begin{figure}
\includegraphics[width=7.5cm]{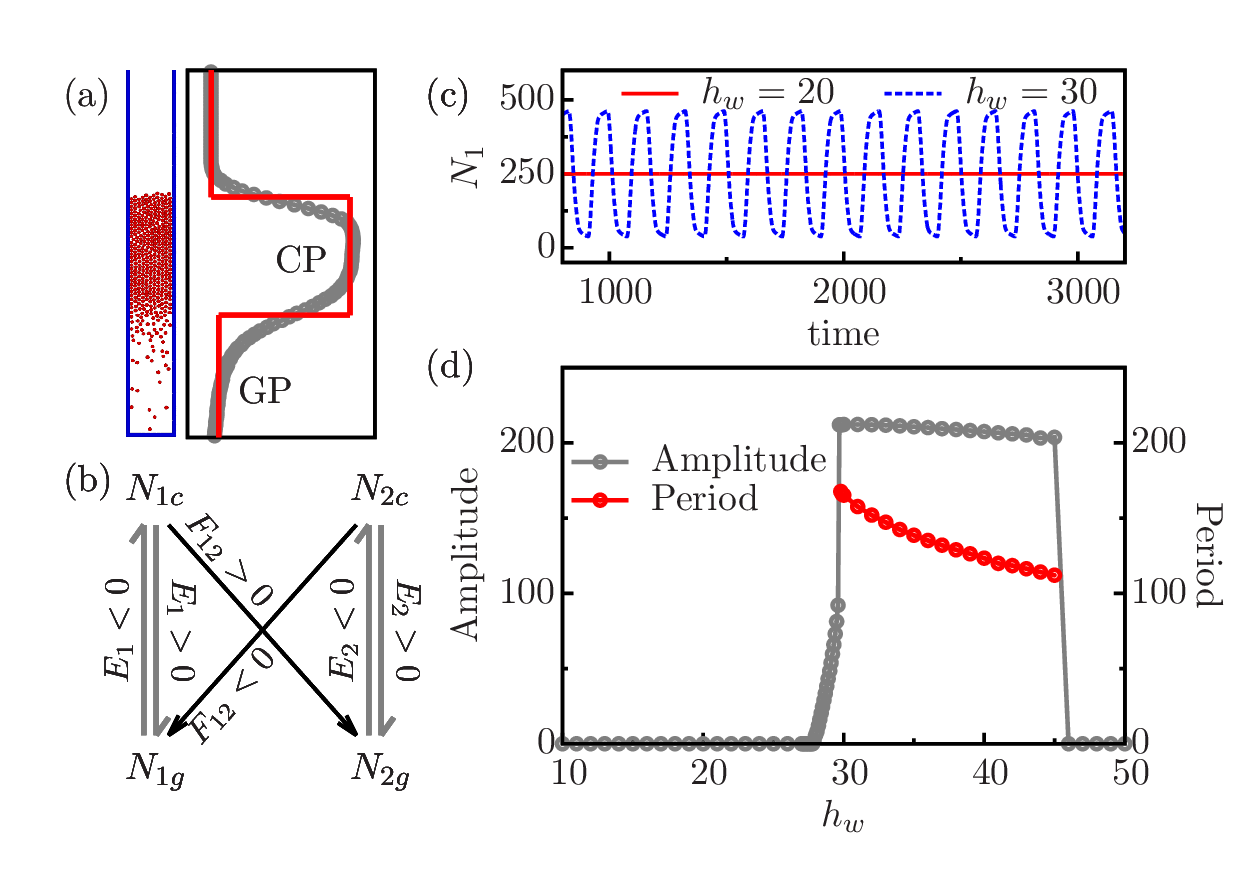}
\caption{
\label{fig5} Theoretical model and numerical solutions for $N=500$ and $v_b=24$: (a) the simplified density profile (red line) for a granular column; (b) the net flows between phases considered in our model; (c) $N_1(t)$ for the steady state at $h_w=20$ and the oscillatory state at $h_w=30$; (d) oscillation amplitude and period of $N_1(t)$ for different $h_w$.
}
\end{figure}

To confirm the theoretical feasibility of the structural matching concept and to clarify the critical role played by the density gradient, we propose a minimal model for the oscillation. We simplify the density profile $n_i(y)$ of any column $i$ by assuming that both GP and CP, respectively with a population of $N_{ig}$ and $N_{ic}=N_i-N_{ig}$, are homogeneous [Fig. \ref{fig5}(a)]. Both the inter-column and the intra-column (between GP and CP) flows of beads need to be considered. First, the outflow flux from column $i$ can be measured as $F_{i} =\lambda \int_0^{h_w} n_i(y)dy$ \cite{Eggers1999,Mikkelsen2004}, where $\lambda$ is a $v_b$ dependent parameter. The net flux between the columns, $F_{12} =-F_{21}=F_1-F_2$, should mainly describe the transfer of beads from CP in one column to GP in the other column, as shown in Fig. \ref{fig4}. To emphasize this point, we further assume constant number densities, $n_g$ and $n_c$, respectively for GP and CP. Thus between the columns, only CP-to-GP flows are allowed due to the density difference, as illustrated in Fig. \ref{fig5}(b). Employing the constant-density assumption, $F_{i}$ can be easily calculated \cite{mono-sm}. Second, we simply describe the flow between GP and CP in column $i$ with a rate $E_{i}=-\beta (N_{ig}-N^s_{ig})$, where superscript $s$ represents the steady state of the column in the uncoupled case, and $\beta$ is another $v_b$ dependent parameter. As shown in Fig. \ref{fig5}(b), $E_i>0$ indicates evaporation and $E_i<0$ describes condensation. We assume that $N^s_{ig}(N_i)=N_i e^{-\alpha N_i}$ \cite{arg-ngs}, with a fitting parameter $\alpha$. Finally, the dynamics of the system can be described by the following equations:
\begin{equation}
\dot{N}_{ig}=E_i + F_{ji} \mathbf{H}(F_{ji}),\;\; \dot{N}_{i} = \dot{N}_{ic}+\dot{N}_{ig} = F_{ji},  \label{eqsdyn}
\end{equation}
where the overdot denotes the derivative with respect to time, $\mathbf{H}(\cdot)$ is the Heaviside step function, $i$ runs from 1 to 2, and $j=3-i$.

When $h_w = 0$, the coupling term $F_{ij}=0$. For an initial state with $N_1=N_2=N/2$, the system represented by Eqs. \eqref{eqsdyn} will obviously stay in an EPSS with fixed $N_{ig}=N_{ig}^s(N/2)\equiv n_gh_{inv}$. Around the EPSS, $\partial F_{12}/\partial N_{ig} \propto S \equiv \lambda(n_c/n_g-1)\mathbf{H}(h_w-h_{inv})$. Linear stability analysis \cite{mono-sm} shows that a Hopf bifurcation with respect to $\beta$ exists for $S>0$ but not for $S=0$. Though $\beta$ cannot be varied in our case, the system is possibly already in the oscillatory regime ($\beta<S$) once the Hopf bifurcation is switched on by $S$. Thus the triggering dynamics of the oscillation can be understood as a switchable two-parameter Hopf bifurcation, and the switch $S$ gives a structural matching boundary $h_{LO}=h_{inv}$ for the oscillation. Moreover, $S\propto(n_c-n_g)$, which also indicates the density gradient around the height of $h_{inv}$, should be large enough ($S>\beta$) to trigger the oscillation. Since neither $h_w$ nor $S$ is the Hopf bifurcation parameter, it is also explained why no obvious Hopf-like behaviors in the amplitude and period [Fig. \ref{fig2}(e)-(f)] are found near the triggering point in the simulation results.

To numerically solve the model, a close packing density $n_c=2/\sqrt{3}W$ for CP and $n_g=0.2W$ for GP are adopted according to the simulation results. We keep $N$ and $v_b$ fixed, say $N=500,v_b=24$. Then, $\alpha \approx 0.006$ can be obtained by fitting the data from the simulation of isolated columns, and this gives $h_{inv}\approx27.9$. We choose $\beta=0.05,\;\lambda=0.20$ to recover similar dynamics to the simulation results. Then Eqs. \eqref{eqsdyn} can be solved with the Runge-Kutta method. A stable fixed-point solution corresponding to the EPSS exists for $S=0$ ($h_w<h_{inv}$), and loses its stability to give way to an oscillation when $S$ becomes positive ($h_w>h_{inv}$), as shown in Fig. \ref{fig5}(c). Our model also sets an upper boundary $h_{HI}\approx 45$ for the oscillation. In the oscillatory regime $h_{LO}<h_w<h_{HI}$, similar behaviors in the oscillation amplitude and period to the simulation results [Fig. \ref{fig2}(e)-(f)] are found, as shown in Fig. \ref{fig5}(d). Hence almost all the characteristics of the oscillation are recovered with this simple model.

In conclusion, a coupling-induced giant oscillation is discovered for the first time in a simple monodisperse granular system, which indicates that high-energy spatial patterns like granular Leidenfrost states may be spontaneously converted into temporal patterns in a nonequilibrium system. The triggering mechanism of the oscillation is confirmed and a switchable Hopf bifurcation is identified by our minimal model. Controlled by the switch parameter, the dynamics of the system differs significantly from a typical Hopf bifurcation.

Furthermore, the oscillation is robust in simulation for different $e$ or inelastic collision models, or for a reasonable range of $W$ \cite{arg-unpub}. Instead of the underlying mechanism of the clustering behavior, the density structure proves to be critical to the oscillation, as also evidenced by our model. A previous study on a similar system \cite{Brey2001} has reported that, in the absence of external fields, clustering behaviors due to dissipation only lead to asymmetric steady states. In such a circumstance, the density structure barely stores any potential energy and cannot provide an efficient feedback mechanism in the coupled dynamics. Thus no oscillation can be observed. Similar to the phenomena of granular Maxwell demon \cite{Brey2007} and the bidisperse granular clock \cite{Hussain2012}, our system can be extended to the case of three or more coupled columns, in which similar oscillations are observed \cite{arg-unpub}. Due to the simplicity and extensibility of our system, the above results may help in understanding some complex biological oscillations. Our further study will focus on the experimental observation of such a monodisperse granular clock.
\begin{acknowledgments}
We thank Dr. Yinchang Li for his preliminary work on the simulation. R. L. thanks Prof. C. K. Chan and P. Y. Lai for useful discussions. This work is supported by National Natural Science Foundation of China (Grant No. 11404378 and 11474326), the MOST 973 Program (Grant No. 2015CB856800), and the Chinese Academy of Sciences ``Strategic Priority Research Program SJ-10'' (Grant No. XDA04020200).
\end{acknowledgments}

\end{document}